\documentclass[aps,prd,preprint,superscriptaddress,nofootinbib,tightenlines]{revtex4-1}
\usepackage{color}
\usepackage{amsmath}
\usepackage{graphicx}
\usepackage{latexsym}
\begin{document}

\preprint{RESCEU-6/18, RUP-18-14}

\title{Self-anisotropizing inflationary universe in Horndeski theory and beyond}

\author{Hiroaki W. H. Tahara}
\affiliation{Research Center for the Early Universe (RESCEU), Graduate School of Science,
The University of Tokyo, Tokyo 113-0033, Japan}
\affiliation{Department of Physics, Graduate School of Science,
The University of Tokyo, Tokyo 113-0033, Japan}

\author{Sakine Nishi}
\affiliation{Department of Physics, Rikkyo University, Toshima, Tokyo 171-8501, Japan}

\author{Tsutomu Kobayashi}
\affiliation{Department of Physics, Rikkyo University, Toshima, Tokyo 171-8501, Japan}

\author{Jun'ichi Yokoyama}
\affiliation{Research Center for the Early Universe (RESCEU), Graduate School of Science,
The University of Tokyo, Tokyo 113-0033, Japan}
\affiliation{Department of Physics, Graduate School of Science,
The University of Tokyo, Tokyo 113-0033, Japan}
\affiliation{Kavli Institute for the Physics and Mathematics of the Universe (Kavli IPMU), WPI, UTIAS,
The University of Tokyo, Kashiwa, Chiba 277-8568, Japan}

\begin{abstract}
As opposed to Wald's cosmic no-hair theorem in general relativity,
it is shown that the Horndeski theory (and its generalization)
admits anisotropic inflationary attractors
if the Lagrangian
depends cubically on the second derivatives of the scalar field.
We dub such a solution as a self-anisotropizing inflationary universe
because anisotropic inflation can occur without introducing
any anisotropic matter fields such as a vector field.
As a concrete example of self-anisotropization
we present the dynamics of a Bianchi type-I universe
in the Horndeski theory.
\end{abstract}

\maketitle

\section{Introduction}
Inflation in the early universe \cite{Starobinsky:1980te,Sato:1980yn,Guth:1980zm,Sato:2015dga}
solves a number of fundamental problems in cosmology, such as horizon, flatness, monopole and the origin-of-structure problems.
Its basic predictions have been confirmed by a number of observations of the cosmic microwave background (CMB)
and large-scale structures.
The simplest single-field inflation paradigm fits, in a sense, observations too well, so that it is difficult to single out the correct field theory model of inflation.
In this context, much work has been done in search for anomalies in observations.
One possibility among them is statistical anisotropy of spectrum of primordial perturbations.
If established, we need an inflation model to realize anisotropic expansion.
In this regard it has been known that inflation driven by a scalar potential make the observable universe fully isotropic \cite{Jensen:1986vy,Turner:1986gj}.
Hence previous models of potential-driven anisotropic inflation inevitably include a vector field \cite{Ford:1989me,Watanabe:2009ct,Do:2017qyd,Adshead:2018emn}.

Here we wish to show that these observations are true only in the Einstein gravity, and that a class of modified gravity with a scalar field can
realize anisotropic (inflationary) solution without introducing any vector fields
nor higher-order curvature terms as an effective anisotropic stress source \cite{Barrow:2005qv,Barrow:2006xb,Barrow:2009gx}.
Specifically we consider Horndeski theory \cite{Horndeski:1974wa} or the generalized Galileon \cite{Deffayet:2011gz} to show that
the quintic galileon term $\mathcal{L}_5$ in generalized G-inflation \cite{Kobayashi:2011nu} plays an essential role to realize anisotropic inflation.
Such terms are known to emerge after Kaluza-Klein compactification of higher-dimensional Lovelock gravity \cite{VanAcoleyen:2011mj} on one hand
but its magnitude has been severely constrained \cite{Baker:2017hug,Creminelli:2017sry,Sakstein:2017xjx,Ezquiaga:2017ekz}
now on the other hand by the simultaneous discovery of the gravitational-wave event GW170817 \cite{TheLIGOScientific:2017qsa}
of binary neutron star coalescence and the associated gamma-ray burst GRB170817A \cite{Goldstein:2017mmi},
which shows that the relative deviation of the speed of gravitational waves from light velocity is at most $10^{-15}$.
Such an observational constraint on $\mathcal{L}_5$, however, applies only in the low-redshift universe,
and it may well evolve nontrivially in the high energy regime in the early universe.

The present paper is organized as follows.
In Section \ref{Horndeskitheory}, we review the covariant form and ADM form of Horndeski theory and its equation of evolution.
In Section \ref{newsolution}, by using the trace-free part of the equation of evolution, we show there are emerged solutions which describe
expanding universes with nonvanishing anisotropies.
In Section \ref{Bianchimodel}, we apply the solution to the Bianchi-I model without matter,
which is the simplest homogeneous anisotropic model.
In Section \ref{discussion}, we disscuss the nature of the solution and cosmological application,
and we conclude in Section \ref{conclusion}.

\section{Horndeski theory and beyond}\label{Horndeskitheory}
The
Horndeski theory \cite{Horndeski:1974wa} describes the most general couplings between a scalar field $\phi$
and the metric $g_{\mu\nu}$
which yield second-order field equations.
It was rediscovered in \cite{Deffayet:2011gz} in the context of generalized Galileons
and their equivalence was proved in \cite{Kobayashi:2011nu}.

This theory is characarized by four arbitrary functions, $G_2$, $G_3$, $G_4$ and $G_5$,
of $\phi$ and its canonical kinetic function $X\equiv-\partial_\mu\phi \partial^\mu\phi/2$ as
\begin{eqnarray}
S&=&\int \mathrm{d}^4x \sqrt{-g} \sum_{i=2}^{5} \tilde{\mathcal{L}}_i, \label{Horndeski}
\\
\tilde{\mathcal{L}}_2&=&G_2(\phi,X),\\
\tilde{\mathcal{L}}_3&=&-G_3(\phi,X)\Box\phi,\\
\tilde{\mathcal{L}}_4&=&G_4(\phi,X){\cal R}
+G_{4X}[(\Box\phi)^2-(\nabla_\mu \nabla_\nu \phi)^2],\\
\tilde{\mathcal{L}}_5&=&G_5(\phi,X){\cal G}_{\mu\nu}\nabla^\mu \nabla^\nu \phi
-\frac{1}{6}G_{5X}[(\Box\phi)^3-3(\Box\phi)(\nabla_\mu \nabla_\nu \phi)^2
+2(\nabla_\mu \nabla_\nu \phi)^3],
\end{eqnarray}
where ${\cal R}$
is the Ricci scalar, ${\cal G}_{\mu\nu}$ is the Einstein tensor,
$(\nabla_\mu\nabla_\nu\phi)^2=\nabla_\mu\nabla_\nu\phi\nabla^\mu\nabla^\nu\phi$,
$(\nabla_\mu\nabla_\nu\phi)^3=\nabla_\mu\nabla_\nu\phi\nabla^\nu\nabla^\lambda%
\phi\nabla_\lambda\nabla^\mu\phi$,
and $G_{iX}=\partial G_i/\partial X$.

The action descibed by the ADM variables
is more useful to study anisotropic cosmological solutions
than the covariant form \eqref{Horndeski}.
The metric is given by
\begin{eqnarray}
ds^2 = -N^2 dt^2 + g_{ij} (dx^i+N^i dt) (dx^j+N^j dt) .
\end{eqnarray}
We take the unitary gauge, $\phi=\phi(t)$, and
then $X$ is given by $X=\dot\phi^2/2N^2$ with $N$ being the lapse function.
If $\phi$ is a monotonic function of $t$,
this is a very convenient gauge and we can use $(t, N)$
instead of $(\phi,X)$ to express the action.
Then, the theory is described only in terms of
$t$ and
geometrical quantities as
\begin{eqnarray}
S&=&\int \mathrm{d}t \mathrm{d}^3x N \sqrt{g} \sum_{i=2}^{5} \mathcal{L}_i ,\label{ADM}\\
\mathcal{L}_2&=&A_2(t,N),\label{admL2}\\
\mathcal{L}_3&=&A_3(t,N){K},\\
\mathcal{L}_4&=&A_4(t,N)\left({K}^2-{K}^i_j{K}^j_i\right)+B_4(t,N){R},\\
\mathcal{L}_5&=&A_5(t,N)\left({K}^3-3{K}{K}^i_j{K}^j_i+2{K}^i_j{K}^j_k{K}^k_i\right)
+B_5(t,N)\left({R}_{ij}-\frac{1}{2}g_{ij}{R}\right){K}^{ij} .\label{admL5}
\end{eqnarray}
$K^i_j$ and $R_{ij}$ are the extrinsic and intrinsic curvature
of constant $t$ (constant $\phi$) hypersurfaces.
The functions $A_i, B_i$ and $G_i$
are related with each other as follows:
\begin{eqnarray}
A_2(t,N) &=& G_2(\phi,X) - \sqrt{X} \int \frac{G_{3\phi}(\phi,X)}{\sqrt{X}} dX, \\
A_3(t,N) &=& \int \sqrt{2X} G_{3X}(\phi,X)  dX - 2 \sqrt{2X} G_{4\phi}(\phi,X), \\
A_4(t,N) &=& - G_4(\phi,X)  + 2X G_{4X}(\phi,X)  - XG_{5\phi}(\phi,X) ,\label{defA4} \\
A_5(t,N) &=& \frac{1}{6} (2X)^{3/2} G_{5X}(\phi,X) , \\
B_4(t,N) &=& G_4(\phi,X)  - \frac{\sqrt{X}}{2} \int \frac{G_{5\phi}(\phi,X) }{\sqrt{X}} dX, \\
B_5(t,N) &=& - \int \sqrt{2X} G_{5X}(\phi,X)  dX\label{defB5},
\end{eqnarray}
where we identify $X=\dot\phi^2(t)/2N^2$.
As seen below, among those terms the most crucial ones in this paper are
the terms cubic in the extrinsic curvature.
In the covariant language they come from $\tilde{{\cal L}}_5$
which depends cubically on the second derivatives of the scalar field.

In the Horndeski theory, $(A_4,A_5)$
and $(B_4,B_5)$ are not independent,
as is clear from
Eqs. \eqref{defA4}--\eqref{defB5} and also from
the fact that we originally have four free functions in the action.
However, this point turns out to be not essential in the following discussion.
The most important ingredient here is the cubic (or higher) order terms
in the extrinsic curvature.
This allows us to start from the ADM Lagrangians \eqref{admL2}--\eqref{admL5}
and consider all $A_i$'s and $B_i$'s to be independent free functions,
which amounts to employing the so-called ``beyond Horndeski'' theory~\cite{Gleyzes:2014dya}, 
although $A_i$'s and $B_i$'s may have to satisfy degeneracy conditions to avoid an extra dangerous degree of freedom 
\cite{Langlois:2015cwa,Crisostomi:2016tcp} (see, however, \cite{DeFelice:2018ewo}).
The following discussion can thus apply not only to the Horndeski theory
but also to beyond Horndeski theory.

In addition to the action for
the gravitational sector described above,
we include the action for matter minimally coupled to gravity, $S_{\rm m}$.
By the use of the residual gauge degrees of freedom one can
further impose $N^i=0$.
Then, we obtain the evolution equations from \eqref{ADM} as
\begin{eqnarray}
T^i_j=
&&\frac{1}{ N\sqrt{g} } \partial_t
\left\{
\sqrt{g}
\left\{
A_3 \delta^i_j
+ 2 A_4(K\delta^i_j - K^i_j)
+ 3 A_5 [(K^2-K^k_l K^l_k)\delta^i_j - 2 (KK^i_j - K^i_k K^k_j)]
\right\}
\right\}
\nonumber\\
&&
- \delta^i_j \mathcal{L}_A
+ \left(2 B_4 + \frac{ \partial_t B_5}{N} \right)\left(
 R^i_j-\frac{1}{2}\delta_i^jR\right)
+\Phi^i_j ,\label{Einstein}
\end{eqnarray}
where $T_{ij}$ is the stress-energy tensor calculated from the matter action $S_\mathrm{m}$,
\begin{eqnarray}
T_{ij} = - \frac{2}{N\sqrt{g}} \frac{\delta S_\mathrm{m}}{\delta g^{ij}} ,
\end{eqnarray}
and
$\mathcal{L}_A$ is the kinetic part of
the Lagrangian,
\begin{eqnarray}
\mathcal{L}_A = A_2 + A_3 K + A_4 (K^2-K^i_j K^j_i) + A_5 (K^3 - 3 K K^i_j K^j_i + 2K^i_j K^j_k K^k_i) .
\end{eqnarray}
We have collected the terms that vanish
if the lapse function is homogeneous, $N(t,\Vec{x})=N(t)$,
and written
\begin{eqnarray}
\Phi_{ij}
&=&
\frac{2}{N}[\nabla^2(NB_4)g_{ij} - \nabla_i \nabla_j (NB_4)]
\nonumber\\
&&+
g_{ij} K^{lm} \nabla_l \nabla_m B_5
+ K \nabla_k \nabla_j B_5
- 2 K^l_{(i} \nabla_{j)} \nabla_l B_5
+ K_{ij} \nabla^2 B_5
- g_{ij} K\nabla^2B_5
\nonumber\\
&&
+ \frac{2}{N}  \left[
g_{ij}\nabla_l (NK^{lm})\nabla_m B_5
+ \nabla_{(i}(NK)\nabla_{j)}B_5
- \nabla_l (NK^l_{(i}) \nabla_{j)} B_5
\right.
\nonumber\\
&&~~~~~~~~~~~~
\left.
- \nabla_{(i} (NK^l_{j)}) \nabla_l B_5
+ \nabla_l (NK_{ij}) \nabla^l B_5
- g_{ij} \nabla_l (NK) \nabla^l B_5
\right] .
\end{eqnarray}
The Hamiltonian constraint is given by
\begin{eqnarray}
&&\partial_N (NA_2) + N \partial_N A_3 K + N^2 \partial_N (N^{-1} A_4)(K^2 - K^i_j K^j_i) + \partial_N (NB_4) R \nonumber\\
&&+ N^3 \partial_N (N^{-2}A_5)(K^3 - 3 K K^i_j K^j_i + 2K^i_j K^j_k K^k_i) + N \partial_N B_5 \left(R_{ij} K^{ij}-\frac{1}{2}RK\right)
 + \frac{1}{\sqrt{g}}\frac{\delta S_\mathrm{m} }{\delta N} = 0.
 \nonumber\\
\label{Hamiltonian}
\end{eqnarray}
In the following we will not use the momentum constraint equations.

\section{Self-anisotropizing inflationary solutions}\label{newsolution}

We now show that even without any anisotropic matter sources
the universe can exhibit anisotropic inflationary expansion as an atractor solution
in the Horndeski theory.

Since we consider Bianchi cosmology, we may set $N^i=0$.
Thanks to the homogeneity, $\Phi_{ij}$ in the evolution
equation \eqref{Einstein} vanishes.
To study anisotropic cosmological models it is convenient to
decompose the extrinsic curvature $K_{ij}$ into its trace $K$ and trace-free part $\Sigma_{ij}$
as
\begin{eqnarray}
K_{ij} = \frac{1}{3} K g_{ij} + \Sigma_{ij} ,
\end{eqnarray}
with $g^{ij} \Sigma_{ij}=0$.
The trace and trace-free parts of the evolution equation \eqref{Einstein} read,
respectively,
\begin{eqnarray}
\frac{1}{ N\sqrt{g} } \partial_t
[
\sqrt{g}
(
3 A_3 + 4 A_4 K + A_5 (2 K^2- 3\Sigma^i_j\Sigma^j_i )
)
]
- 3 \mathcal{L}_A
- \left( B_4 + \frac{\partial_t B_5}{2N} \right) R
=T^i_i . \label{tracepart}
\end{eqnarray}
and
\begin{eqnarray}
\frac{2}{N\sqrt{g}} \partial_t
\left[
\sqrt{g} (-A_4 \Sigma^i_j - A_5 K \Sigma^i_j + 3 A_5 \{ \Sigma^i_k \Sigma^k_j \}_\mathrm{TF})
\right]
+
\left(2 B_4 + \frac{\partial_t B_5}{N} \right) \{ R^i_j \}_\mathrm{TF}
=
\{ T^i_j \}_\mathrm{TF} ,
\nonumber\\
 \label{tracefree}
\end{eqnarray}
where $\{ X^i_j \}_\mathrm{TF}$ stands for the trace-free part of a tensor $X^i_j$,
\begin{eqnarray}
\{ X^i_j \}_\mathrm{TF} = X^i_j - \frac{1}{3} X^k_k \delta^i_j .
\end{eqnarray}

Let us look for slow-roll inflationary solutions
in which $\sqrt{g}$ exponentially increases,
while other functions remain either nearly constant or exponentially decrease.
First,
we focus on Eq. \eqref{tracefree}, assuming that
 the energy-momentum tensor consists of isotropic matter
and hence $\{ T^i_j \}_\mathrm{TF}$ vanishes.
If the spatial curvature
$R^i_j$ decreases exponentially,
the first term also decreases in the same way.
As a result, we find, asymptotically,
\begin{eqnarray}
-A_4 \Sigma^i_j - A_5 K \Sigma^i_j + 3 A_5 ( \Sigma^i_k \Sigma^k_j - \frac{1}{3}\Sigma^k_l \Sigma^l_k \delta^i_j) = 0 .
\label{attractor}
\end{eqnarray}
A trivial solution of Eq.~\eqref{attractor} is that all components of $\Sigma^i_j$ vanish.
This solution corresponds to the isotropic attractor
which we see in the conventional inflation models.
The presence of the quadratic terms
in $\Sigma^i_j$ due to nonvanishing $A_5$ yields nontrivial solutions with
$\Sigma^i_j\neq 0$ as well,
which represent an expanding universe retaining
finite anisotropies.
We dub this anisotropic attractors as {\em self-anisotropizing} inflationary solutions,
as this is {\em not} caused by an anisotropic energy-momentum tensor.
We will demonstrate in the next section that such solutions do exist
in the case of Bianchi type-I cosmology.

The self-anisotropizing attractors are distinct from the previous
anisotropic inflationary solutions, because the anisotropic expansion of
the previous scenarios are supported by some anisotropic energy-momentum source
such as a vector field coupled with an inflaton field \cite{Watanabe:2009ct}.
Such scenarios produce background anisotropies
$\Sigma^i_j/H \approx \{T^i_j \}_\mathrm{TF}/(6 A_4 H^2) = (8\pi G/3H^2) \{T^i_j \}_\mathrm{TF} $,
where $H$ is the Hubble parameter.
The trace-free part of the energy-momentum tensor, $\{T^i_j \}_\mathrm{TF}$,
just displaces the terminal point from the isotropic one.

By contrast, here
the self-anisotropizing inflationary solution
is realized by the terms quadratic in $\Sigma^i_j$ in Eq. \eqref{attractor},
which is a consequence of modification of gravity.
The magnitude of produced background anisotropies is
estimated from \eqref{attractor}
as $\Sigma^i_j/H\sim (A_4 + 3 H A_5)/3 H A_5$.
We require neither an anisotropic energy-momentum tensor
nor any fields other than the scalar $\phi$ built in the Horndeski theory.
In this sense, the emerged anisotropic terminal points should be distinguished from
those of previous anisotropic inflation models.

Let us evaluate the eigenvalues of the nontrivial solutions of $\Sigma^i_j$ for
given values of $A_4,~A_5$ and $K$.
We can prove that the root $\Sigma$ of matrix equation \eqref{attractor}
has two different eigenvalues at most as follows.
First we define a polynomial $p(x)$ by substituting a real variable $x$ for $\Sigma$ in the left side of \eqref{attractor} as
\begin{eqnarray}
p(x) = -A_4 x - A_5 K x + 3 A_5 \left( x^2 -\frac{1}{3} \mathrm{tr}\!\left(\Sigma^2\right) \right) , \label{minimal}
\end{eqnarray}
where the remaining $\Sigma$ in the trace is a root of \eqref{attractor}.
$p(\Sigma)=0$ obviously follows from \eqref{attractor} and \eqref{minimal},
and so $p(x)$ can be divided by the minimal polynomial $\phi_\Sigma(x)$ of $\Sigma$.
In linear algebra, it is well-known that if $\lambda$ is an eigenvalue of matrix $\Sigma$ then $\lambda$ is a root of $\phi_\Sigma(x)=0$.
Therefore, the eigenvalue $\lambda$ is also a root of $p(x)=0$.
Since $p(x)$ is a quadratic polynomial of $x$, the number of different roots is equal to or less than two.
This is the proof that $\Sigma$ has two different eigenvalues, $\lambda_1$ and $\lambda_2$ at most.
It induces that, \textit{e.g.}, anisotropic attractors in Bianchi type-I model has axial symmetry in the order of background,
which we show in Section \ref{Bianchimodel}.
As one can see from \eqref{minimal}, the different eigenvalues $\lambda_1$ and $\lambda_2$ satisfy
\begin{eqnarray}
\lambda_1 + \lambda_2
=
\frac{A_4+A_5 K}{3A_5}.
\end{eqnarray}
Being a three dimensional tensor, $\Sigma$ has three eigenvalues.
Without loss of generality, we set them as $\lambda_1$, $\lambda_1$ and $\lambda_2$, respectively. They also satisfy
\begin{eqnarray}
2 \lambda_1 + \lambda_2 = 0,
\end{eqnarray}
because $\Sigma$ is trace-free. Therefore we have
\begin{eqnarray}
\lambda_1 = -\frac{A_4+A_5K}{3A_5}
,~~~~
\lambda_2 = \frac{2(A_4+A_5K)}{3A_5}.
\label{eigenvalues}
\end{eqnarray}

So far we have focused on the evolution equation for $\Sigma^i_j$ \eqref{tracefree}
and its nontrivial solution under the assumption that
the spatial volume element $\sqrt{g}$ increases exponentially
and the spatial curvature $R^i_j$ decreases accordingly.
To determine all the components of the metric, we need to
solve the Hamiltonian constraint \eqref{Hamiltonian} and
the trace part of the evolution equations \eqref{tracepart} consistently.
On the anisotropic attractor where
$\Sigma^i_j$'s eigenvalues are given by \eqref{eigenvalues},
the rest of the field equations
\eqref{Hamiltonian} and \eqref{tracepart} reduce to
\begin{align}
\partial_N \left[ N \left( A_2 - \frac{2 A_4^3}{9 A_5^2} \right) \right]
+ N \partial_N \left( A_3 -\frac{2 A_4^2}{3 A_5} \right) K
&=
- \frac{1}{\sqrt{g}}\frac{\delta S_\mathrm{m} }{\delta N}
\label{eq:at1}
,
\\
\frac{1}{ N } \frac{d}{dt}
\left( A_3 -\frac{2 A_4^2}{3 A_5} \right)
- \left( A_2 - \frac{2 A_4^3}{9 A_5^2} \right)
&=\frac{1}{3} T^i_i,\label{eq:at2}
\end{align}
respectively.
These two equations can be used to determine $K=K(t)$ and $N=N(t)$.

Let us ignore the matter field $S_{\rm m}$ for the moment
and consider a theory with (approximate) shift symmetry.
In this case, $A_i$'s depend only on $\dot\phi/N$
and from Eq. \eqref{eq:at2}
one obtains a solution $N\simeq {\rm const}\times \dot\phi$
satisfying $F(\dot\phi/N)\equiv A_2- 2A_4^3/9A_5^2\simeq 0$. Equation \eqref{eq:at1} is then
solved to give $K\simeq -\partial_N(A_3-2A_4^2/3A_5)/\partial_N(A_2-2A_4^3/9A_5^2)%
\simeq\;$const. One thus obtains an inflating solution with nonvanishing anisotropies.

\section{Vacuum Bianchi type-I model}\label{Bianchimodel}
\subsection{Evolution toward attractors}
To be more explicit, let us consider
the Bianchi type-I model, which is the simplest homogeneous anisotropic model
and hence helps us to understand what nonvanishing $A_5$ causes.

Once we diagonalize the spatial metric and its time derivative, off-diagonal components are not generated in this model,
so that we can express the metric in the Kasner-type form as
\begin{eqnarray}
ds^2&=&-N^2(t)dt^2
+a^2(t)
\left[
e^{2(\beta_+(t)+\sqrt{3}\beta_-(t))}dx^2 +
e^{2(\beta_+(t)-\sqrt{3}\beta_-(t))}dy^2 +
e^{-4\beta_+(t)}dz^2
\right] ,
\label{metric}
\end{eqnarray}
where $a(t)$ is a scale factor and $\beta_\pm(t)$ show the
differences between the expantion rates in different directions.
Substituting the metric \eqref{metric} in the ADM form of the action \eqref{ADM},
we obtain
\begin{eqnarray}
 S &=&
 \int dt d^3x~
N a^{3}
 \left[
 A_2 + 3 H A_3 + 6 A_4 (H^2-\sigma_+^2-\sigma_-^2)
 \right.
 \left.
 + 6A_5(H^3-3H(\sigma_+^2+\sigma_-^2)+2(3\sigma_+\sigma_-^2 - \sigma_+^3))
 \right],
 \nonumber\\
 \label{action0}
\end{eqnarray}
where we defined the Hubble parameter $H$ and the shear $\sigma_\pm$ as
\begin{eqnarray}
H \equiv \frac{1}{N}\frac{d\ln a}{dt} , ~~~~
\sigma_\pm \equiv \frac{1}{N}\frac{d\beta_\pm}{dt}.
\end{eqnarray}
Using $\sigma_\pm$, the trace-free part of the extrinsic curvature is given by
\begin{eqnarray}
\Sigma^i_j = \mathrm{diag} \left( \sigma_+ + \sqrt{3} \sigma_- , \sigma_+ - \sqrt{3} \sigma_- , -2 \sigma_+ \right).
\label{Sigma}
\end{eqnarray}
Since the spatial Ricci tensor vanishes in the Bianchi type-I model
and consequently Eq. \eqref{action0}
depends on $\beta_\pm$ only through their time derivatives,
the conjugate momenta of $\beta_\pm$ are conserved in time.
The conserved momenta are given by
\begin{eqnarray}
P_{\beta_+}
&=&
a^3
\left[
(A_4+3HA_5)\sigma_+
+3A_5(\sigma_+^2-\sigma_-^2)
\right],\label{Pp}
 \\
P_{\beta_-}
&=&
a^3
\left[
(A_4+3HA_5)\sigma_-
-6A_5\sigma_+ \sigma_-
\right] .
\label{Pm}
\end{eqnarray}
Equivalently,
one can also obtain the same conserved momenta
from the trace-free part of the evolution equations \eqref{tracefree}
by substituting Eq. \eqref{metric}.
It is manifest that as the scale factor $a(t)$ increases,
the expressions inside the square brackets of Eqs. \eqref{Pp} and \eqref{Pm}
decay toward zero as $[\cdots] = P_{\beta_\pm}a^{-3} \to 0$, and
thus $\sigma_+$ and $\sigma_-$ evolve to one of the fixed points.
In the present case, there are four fixed points.
One is the isotropic solution $\sigma_\pm=0$,
whereas the other three are anisotropic attractors.

\begin{figure}[htbp]
  \begin{center}
       \includegraphics[width=10cm]{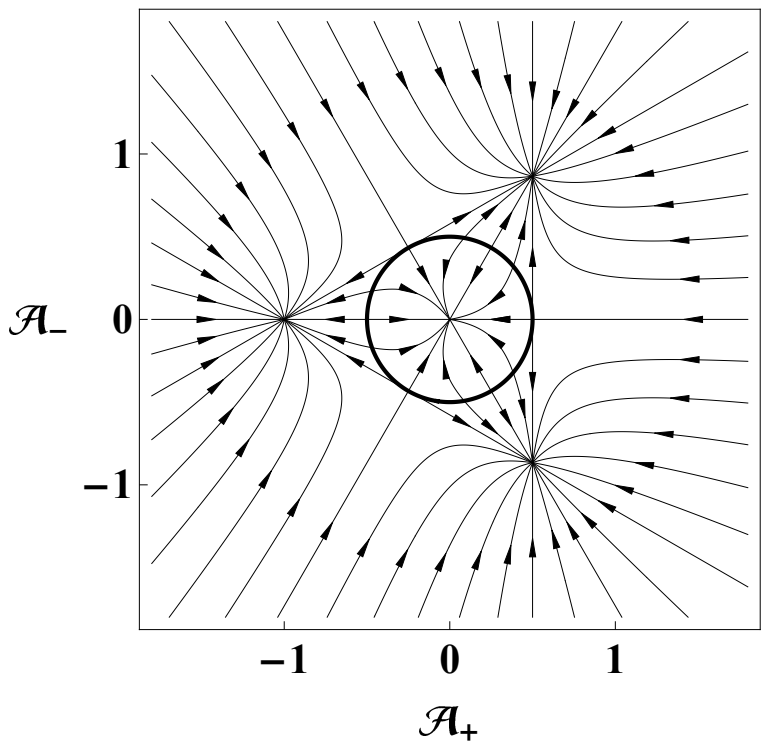}
       \caption{
       Trajectries of the evolution of the normalized shear $({\cal A}_+,{\cal A}_-)$.
       If the initial conditions lie inside the circle given by
       $\mathcal{A}_+^2+\mathcal{A}_-^2=1/4$,
       the universe evolves toward the center, $(\mathcal{A}_+,\mathcal{A}_-)=(0,0)$,
      as $\tau$ increases.
      If the universe starts from outside of the circle,
       it goes to the closest one of the anisotropic fixed points on the vertices,
       $(\mathcal{A}_+,\mathcal{A}_-)=(-1,0),(1/2,\pm \sqrt{3}/2)$, of the triangle as $\tau$ increases.
       }
       \label{attractors}
  \end{center}
\end{figure}

Let us look at the trajectories on the $(\sigma_+,\sigma_-)$ plane of
the phase space.
Given the initial data, the constants $P_{\beta_\pm}$ are fixed.
Then, $\sigma_\pm$ can be expressed in terms of $A_4$, $A_5$, $a$, $H$, and $P_{\beta_\pm}$
by
solving the algebraic equations \eqref{Pp} and \eqref{Pm}.
In order to show the dynamics of the anisotropies in a single figure,
we use the normalized shear ${\cal A}_\pm$ defined as
\begin{eqnarray}
\mathcal{A}_\pm \equiv \frac{ 3A_5 }{ A_4 + 3HA_5 }  \sigma_\pm ,
\end{eqnarray}
instead of $\sigma_+$ and $\sigma_-$. Here we
assumed that $A_4+3HA_5\ne 0$ and $A_5\ne 0$.
It is also convenient to introduce the new time coordinate
$\tau \equiv a^3 (A_4 + 3HA_5)^2 / 3A_5$.
In an expanding universe, $|\tau|$ is an increasing function of $t$
provided that $A_4$, $A_5$, and $H$ depend on $t$ only weakly,
which is a natural assumption during inflation.
With $\tau$ and $\mathcal{A}_\pm$, we can rewrite Eqs. \eqref{Pp} and \eqref{Pm}
simply as
\begin{eqnarray}
P_{\beta_+}
&=&
\tau
\left[
\mathcal{A}_+
+\mathcal{A}_+^2-\mathcal{A}_-^2
\right],\label{nPp}
 \\
P_{\beta_-}
&=&
\tau
\left[
\mathcal{A}_-
-2\mathcal{A}_+ \mathcal{A}_-
\right] . \label{nPm}
\end{eqnarray}
We show trajectories $({\cal A}_+(\tau), {\cal A}_-(\tau))$
for different values of $P_{\beta_\pm}$ in Figure \ref{attractors}.
As stated above,
there are four fixed points in the $(\mathcal{A}_+,\mathcal{A}_-)$ plane:
one isotropic solution, $(0,0)$, and three anisotropic solutions,
$(-1,0)$ and $(1/2,\pm \sqrt{3}/2)$. All of them
are attractors (as long as $|\tau|$ is an increasing function of $t$).

The initial anisotropies determine which attractor the universe approaches.
To see this explcitly, we
differentiate Eqs. \eqref{nPp} and \eqref{nPm} and
get
\begin{eqnarray}
\tau \frac{d \mathcal{A}_+}{d\tau}
=
- \frac{  (2\mathcal{A}_+ - 1)
(\mathcal{A}_+^2+\mathcal{A}_-^2+\mathcal{A}_+)}{4\mathcal{A}_+^2 + 4\mathcal{A}_-^2 - 1 } ,\label{eq:dap}
\\
\tau \frac{d \mathcal{A}_-}{d\tau}
=
- \frac{ \mathcal{A}_- (2\mathcal{A}_+^2 + 2\mathcal{A}_-^2 - 2\mathcal{A}_+ - 1)}{ 4\mathcal{A}_+^2 + 4\mathcal{A}_-^2 - 1 } .\label{eq:dam}
\end{eqnarray}
Equivalently, one may introduce
the polar coordinates $(r(\tau),\theta(\tau))$ defined by ${\cal A_+}=r\cos\theta$
and ${\cal A}_-=r\sin\theta$ and write
\begin{align}
\tau \frac{dr}{d\tau}&=-\frac{r[2r^2+r\cos(3\theta)-1]}{4r^2-1},
\label{eq:dr}
\\
\tau\frac{d\theta}{d\tau}&=\frac{r\sin(3\theta)}{4r^2-1}.\label{eq:dtht}
\end{align}
The denominators vanish on a circle given by
$r^2={\cal A}_+^2+{\cal A}_-^2=(1/2)^2$
(the black circle in Fig.~\ref{attractors}).\footnote{The shear evolution
equations bocome singular on this circle. However,
if we consider the full phase space by
taking into account the trace part of the evolution equation,
we see that this singularity is only apparent.}
The fate of the universe depends on whether the
initial anisotropies are inside this circle or not:
the universe is attracted toward the isotropic solution
at the origin if the initial anisotropies lie inside the circle,
while it goes away from the circle to
the closest one of the anisotorpic attractors
if outside initially.
That is to say, if the universe is sufficiently anisotropic initially,
then it converges to the anisotropic attractor.

The exceptional case is the trajectories with $\theta=0, 2\pi/3, 4\pi/3$.
Those constant values of $\theta$ solve Eq. \eqref{eq:dtht},
while Eq. \eqref{eq:dr} leads to
$r(\tau)=(\sqrt{C/|\tau|+1}-1)/2$, where $C$ is an integration constant.
Therefore, for all initial conditions on $\theta=0, 2\pi/3, 4\pi/3$
the isotropic universe is the attractor.

The structure of Fig.~\ref{attractors} will be more transparent
in terms of the polar coordinates.
Equations \eqref{eq:dr} and \eqref{eq:dtht}
clearly show that there are discrete rotation symmetry $\theta\to \theta + 2\pi/3$
and reflection symmetry across $\theta=0, 2\pi/3$, and $4\pi/3$ axises.
Because of these symmetries only a sixth part of Fig.~\ref{attractors}
is physically independent.

Each of the anisotropic attractors corresponds to an axially symmetric space,
whose symmetry axis is the $x$, $y$ or $z$ axis.
This axial symmetry is closely related to the degeneracy of the eigenvalues of $\Sigma_i^j$
discussed in the previous section.
The discrete rotation symmetry in the $(\mathcal{A}_+,\mathcal{A}_-)$
plane is the manifestation of the fact that
one can always take, say, the $z$ axis as the symmetry axis
without loss of generality by a rotation of the spatial coordinates.

So far we have focused only on the shear evolution equations.
This is sufficient for the purpose of seeing that
the anisotropic fixed points do exist and
for initial anisotropies larger than a certain threshold
they are indeed the attractors.
To determine the precise dynamics of the universe
including the evolution of $H$ and $\phi$, one needs to
solve the full set of the field equations
(the trace and trace-free parts of the evolution equations
as well as the constraint equation) consistently.
In the next subsection we will show a
numerical example obtained by
solving all the equations consistently.

\subsection{Examples}

Let us present some examples which yield self-anisotropizing
Bianchi type-I solutions. The first one is simply given by
\begin{align}
G_2=-V_0,\quad G_3=0,
\quad G_4=\frac{M^2}{2}+g_4X,\quad G_5=g_5X,
\end{align}
where $V_0$, $M$, $g_4$, and $g_5$ are constants.
The corresponding ADM form in the unitary gauge is given by
\begin{align}
&A_2=-V_0,\quad A_3=0,\quad
A_4=-\frac{M^2}{2}+\frac{g_4}{2}\left(\frac{\dot\phi}{N}\right)^2,
\quad A_5=\frac{g_5}{6}\left(\frac{\dot\phi}{N}\right)^3,
\notag \\ &
\quad B_4=\frac{M^2}{2}+\frac{g_4}{2}\left(\frac{\dot\phi}{N}\right)^2,
\quad B_5=-\frac{g_5}{3}\left(\frac{\dot\phi}{N}\right)^3.
\end{align}

\begin{figure}[htbp]
  \begin{center}
       \includegraphics[width=10cm]{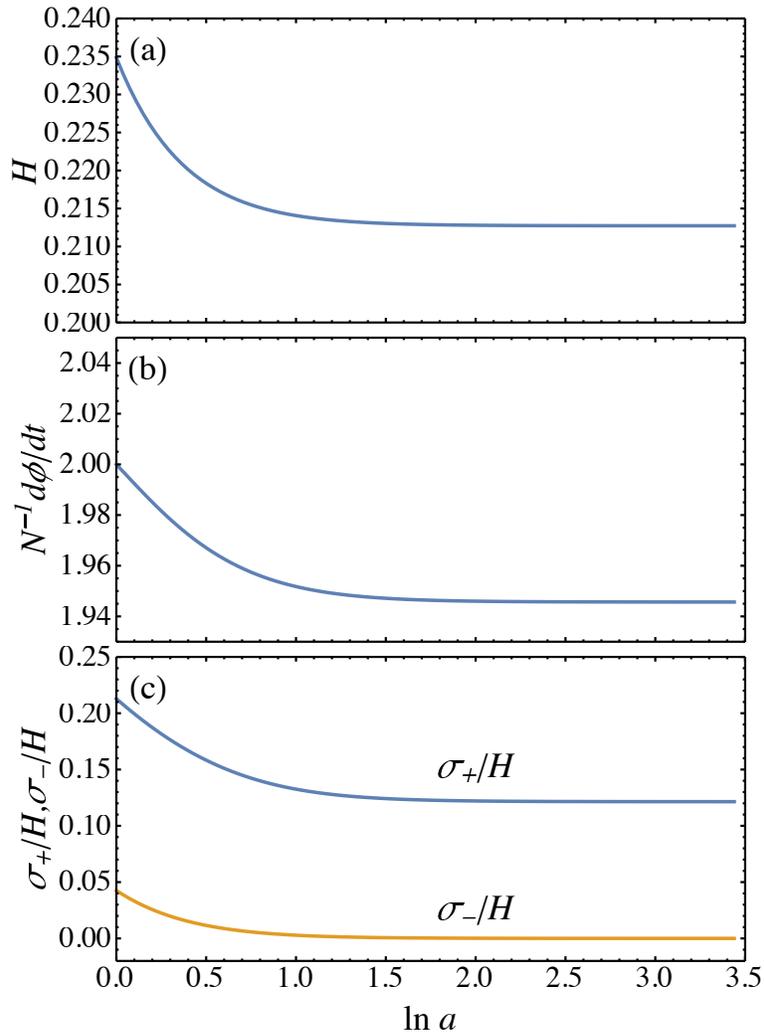}
       \caption{
       Numerical example of a self-anisotrpizing Bianchi type-I universe:
       (a) $H$; (b) $\dot\phi/N$; (c) $\sigma_\pm/H$ as functions of $\ln a$.}
       \label{fig:ex1}
  \end{center}
\end{figure}

Figure~\ref{fig:ex1} shows the evolution of the Hubble parameter,
(the velocity of) the scalar field, and the shear
obtained by solving the dynamical and constraint equations numerically
with a certain initial condition away from the attractors at $a=1$.
The parameters in this toy example are given by
$V_0=0.1$, $M=1$, $g_4=-0.2$, and $g_5=1$.
It can be seen that the universe quicly converges to
the anisoropic inflationary attractor.

Another example with $A_5$ (or, equivalently, $G_{5X}$)
is the Gauss-Bonnet term coupled to a scalar field,
and the total Lagrangian is of the form
\begin{align}
{\cal L}=f(\phi){\cal R}+P(\phi, X)+\xi(\phi)
\left({\cal R}^2-4{\cal R}_{\mu\nu}{\cal R}^{\mu\nu}
+{\cal R}_{\mu\nu\rho\sigma}{\cal R}^{\mu\nu\rho\sigma}\right).
\label{eq:LagGB}
\end{align}
Aspects of
this theory has been studied extensively in the literature.
The Lagrangian can be reproduced by taking the following
Horndeski functions \cite{Kobayashi:2011nu}:
\begin{align}
&
G_2=P+8\xi^{(4)}X^2(3-\ln X),\quad G_3=4\xi^{(3)}X(7-3\ln X),
\notag \\
&G_4=f+4\xi^{(2)}X(2-\ln X),
\quad
G_5=-4\xi^{(1)}\ln X,
\end{align}
where $\xi^{(n)}=d^n\xi/d\phi^n$. Though this looks quite non-trivial,
the corresponding ADM form is very simple:
\begin{align}
A_2=P,\quad  A_3=-2\frac{\dot\phi}{N}\frac{d f}{d\phi},
\quad A_4=-f,\quad A_5=-\frac{4\xi^{(1)}}{3}\frac{\dot\phi}{N},
\quad B_4=f,\quad B_5=8\xi^{(1)}\frac{\dot\phi}{N}.
\end{align}
Even this familiar theory admits self-anisotropizing inflationary solutions.

The theory \eqref{eq:LagGB} possesses a shift symmetry
if $f=\;$const, $P=P(X)$, and $\xi\propto \phi$.
In this case it is easy to find an inflationary solution
with $H=\;$const, $\dot\phi/N=\;$const retaining
the nonvanishing shear
\begin{align}
\frac{\sigma_\pm}{H}\sim \frac{f+4H\xi^{(1)}\dot\phi/N}{H\xi^{(1)}\dot\phi/N}.
\end{align}

\section{Discussion}\label{discussion}

It has been pointed out by Wald that in general relativity, all vacuum Bianchi universes with a positive cosmological constant except type IX evolve toward the isotropic attractor,
which was proven by using the Hamiltonian constraint and the trace of
the Einstein equations \cite{Wald:1983ky}.
In our case, since the Horndeski action dramatically changes both of them,
it must be checked one by one whether a specific model under consideration
evolves toward the isotropic or anisotropic attractor.
We note that the magnitude of
the shear on the anisotropic attractors
diverges when we take the general relativity limit $A_5 \to 0$ keeping $A_4$ constant.
In this limit,
for all initial conditions the isotropic universe is an attractor
(as they are all inside the circle in Fig.~\ref{attractors}),
and thus the standard result of Wald in gerenal relativity is recovered.

Noting that
the background anisotropies of the Bianchi type-I universe can be regarded
as gravitational waves with infinitely long wavelengths,
we point out that
the emergence of anisotropic attractors
is closely related to
the three-point coupling of gravitational waves in the Horndeski theory.
From Eq. (15) of \cite{Gao:2011vs}, one sees that
there are two types of the three-point couplings
of the form $hh\partial^2h$ and $\dot h\dot h\dot h$,
giving rise to local and equilateral non-Gaussianity, respectively.
The former appears even in general relativity as well as in a generic scalar-tensor theory,
while the latter, which obviously comes from $K_{ij}^3$, emerges
only in the class with $A_5\ne 0$ (i.e., $G_{5X}\neq 0$).
The former has spatial derivatives and therefore vanishes in the long-wavelength limit,
whereas the latter has only time derivatives and hence does not vanish
even in the homogeneous limit.

Since ${\cal A_{\pm}}={\cal O}(1)$ on the anisotropic attractors,
the magnitude of the resultant anisotropy is given by
\begin{align}
\frac{\sigma_\pm}{H}\sim \frac{A_4+3HA_5}{3HA_5},
\end{align}
which is typically of ${\cal O}(1)$ or larger.
In theories with $G_{5X}\neq 0$, initial anisotropies must be
smaller than this value in order to realize an isotropic universe
through inflation. Otherwise, the resultant universe would be
unacceptably anisotropic.
Another possiblity is that one has
$A_4+3HA_5\ll 3HA_5$ via fine-tuning,
leaving an observationally viable universe with only tiny anisotropies
on the anisotropic attractor. This would be a very interesting scenario,
but one has to study reheating, cosmological perturbations, and
the stability in detail to see whether such a scenario is indeed viable or not,
which is beyond the scope of the present paper.

\section{Conclusions}\label{conclusion}
We have shown the quintic galileon term proportional $G_{5X}$ or $A_5$ in generalized G-inflation can realize
anisotropic inflationary solution.
On the anisotropic attractor, the Hamiltonian constraint becomes a linear equation for the Hubble parameter
which is strikingly different from the conventional Friedmann equation.

Although our solution generically produces anisotropy of the order of unity or larger,
it can also accommodate much smaller anisotropy
by partially canceling $A_4$ and $3HA_5$.
In order to see if observationally viable anisotropic inflation is possible,
we must calculate perturbations as well as discuss transition to the Friedmann regime with proper reheating,
which will be discussed in future publications.

It is also interesting to study higher dimentional models in this context to show a new compactification mechanism of
extra dimensions in the presence of the highest-order galileon terms in the dimension under consideration.
As we can show that the eigenvalues of the extrinsic curvature tensor take only two distinct values at most
even in higher dimensional models, this may provide a promissing mechanism of compactification or dimensional reduction,
which will also be discussed in a forthcoming paper.

\begin{acknowledgments}
HWHT was supported by the Advanced Leading Graduate Course for Photon Science (ALPS).
  The work of TK was supported by
  MEXT KAKENHI Grant Nos.~JP15H05888, JP16H01102, JP17H06359, JP16K17707,
  and MEXT-Supported Program for the Strategic Research Foundation at Private Universities,
  2014-2018 (S1411024).
The work of JY was supported by JSPS KAKENHI, Grant JP15H02082
and Grant on Innovative Areas JP15H05888.
\end{acknowledgments}

\bibliography{mypaper}

\end{document}